\pdfoutput=1

\documentclass{INCLUDES/llncs}
\usepackage{amsfonts,amsmath}
\usepackage{graphicx}
\usepackage{url}
\usepackage{verbatim} 
\usepackage{listings}
\usepackage{amsmath}
\usepackage{subfig}
\lstnewenvironment{codex}
    {\lstset{}%
      \csname lst@SetFirstLabel\endcsname}
    {\csname lst@SaveFirstLabel\endcsname}
\lstnewenvironment{code}
    {\lstset{}%
      \csname lst@SetFirstLabel\endcsname}
    {\csname lst@SaveFirstLabel\endcsname}
\newtheorem{prop}{Proposition}
\newtheorem{df}{Definition}

\newcommand{\BI}[0]{\begin{itemize}}
\newcommand{\EI}[0]{\end{itemize}}

\newcommand{\BE}[0]{\begin{enumerate}}
\newcommand{\EE}[0]{\end{enumerate}}

\newcommand{\BX}[0]{\begin{codex}}
\newcommand{\EX}[0]{\end{codex}}
\newcommand{\BV}[0]{\begin{verbatim}}
\newcommand{\EV}[0]{\end{verbatim}}

\def \bscale1 {0.25}
\def \bscale {0.25}

\newcommand{\FIG}[4]{
\begin{figure}[htbp]
\centering
{\includegraphics[scale=#3]{./figs/#4}}
\caption{#2}
\label{#1}
\end{figure}
}

\begin{document}

\title{
  Computing with Hereditarily Finite Sequences
}

\author{Paul Tarau}
\institute{
   {Department of Computer Science and Engineering}\\
   {University of North Texas}\\ 
   {\em tarau@cs.unt.edu}\\
}
\maketitle
\date{}

\label{firstpage}

\begin{abstract}
We use Prolog as a flexible meta-language to provide executable
specifications of some fundamental mathematical objects and their
transformations. In the process, 
isomorphisms are unraveled between
natural numbers and combinatorial objects 
(rooted ordered trees representing hereditarily finite sequences
and  rooted ordered binary trees 
representing G\"odel's System {\bf T} types).

This paper focuses on an application
that can be seen as an unexpected ``paradigm shift'':
we provide recursive definitions
showing that the resulting representations
are directly usable to perform symbolically
arbitrary-length integer computations.

Besides the theoretically interesting fact 
of ``breaking the arithmetic/symbolic
barrier'', the arithmetic 
operations performed with symbolic objects
like trees or types 
turn out to be genuinely efficient -- we derive
implementations with asymptotic 
performance comparable to ordinary
bitstring implementations of arbitrary-length
integer arithmetic.

The source code of the paper, organized as a literate Prolog program,
is available at
\url{http://logic.cse.unt.edu/tarau/research/2011/pPAR.pl}

\end{abstract}

\section{Introduction}

This paper exhibits a creative use of logic programming as a modeling tool for
several interesting concepts at the intersection of combinatorics,
formal languages, foundation of mathematics and coding theory.
It builds on the declarative data transformation framework introduced in 
\cite{ppdp09pISO,everything},
where we introduce a methodology to derive bijective mappings between
fundamental data types used in programming languages
(sets, multisets, sequences to graphs, digraphs, DAGs, hypergraphs etc.)

At the same time, with practical uses for arbitrary
size integer arithmetic in mind,
we will focus on keeping the asymptotic complexity of various
operations similar to that of operations
on conventional bitstrings.

Like \cite{ppdp09pISO}, this paper is organized 
as a literate Prolog program. This means that our
``lingua franca'' is logic programming rather than the usual
mathematical notation. 

It has been a long tradition in logic programming to model
 program properties and behaviors in terms of mathematical
 reasoning. We pay it back this time, and model some 
 intriguing mathematical concepts as logic programs. 

The paper is organized as follows.

Section \ref{cons} overviews, following \cite{ppdp09pISO}
a bijection between natural numbers and sequences that is extended in
section \ref{hered}, by recursive application,
to hereditarily finite sequences.
Section \ref{comput} describes a novel way to perform 
arbitrary length arithmetic computations
using multiway tree representations of 
hereditarily finite sequences
and
discusses some potential applications for
implementation of
arithmetic operations with numbers 
that do not fit in computer memory with conventional 
binary encodings.
It is followed by a sketch of similar mechanism in section \ref{systemT} 
for the type language of G\"odel's system {\bf T}.
Section \ref{parlang} introduces a bijection between hereditarily finite
sequences and balanced parenthesis languages providing 
a succinct representation
for them.
Sections \ref{related} and \ref{concl} discuss related work and conclude the
paper.

\section{A bijection between finite sequences and natural numbers}
\label{cons}
Let $\mathbb{N}$ be the set of natural numbers and $[\mathbb{N}]$ the set
of finite sequences of natural numbers (that can also be seen as the set of
functions from an initial segment of  $\mathbb{N}$ to $\mathbb{N}$ - 
or even more generally, as {\em finite functions}).
We will first derive, following \cite{ppdp09pISO} a bijection
$\mathbb{N} \to  [\mathbb{N}]$.

We define the following predicates working on natural numbers:
\begin{code}
cons(X,Y,XY):-X>=0,Y>=0,XY is (1+(Y<<1))<<X.

hd(XY,X):-XY>0,P is XY /\ 1,hd1(P,XY,X).

  hd1(1,_,0).
  hd1(0,XY,X):-Z is XY>>1,hd(Z,H),X is H+1.

tl(XY,Y):-hd(XY,X),Y is XY>>(X+1).

null(0).
\end{code}
After observing that the relations
{\tt cons(X,Y,Z), hd(Z,X), tl(Z,Y)} hold if and only if $Z=2^X(2Y+1)$,
it can be proven by structural induction that:
\begin{prop}
The predicates {\tt
cons/3, hd/2, tl/2, null/1} emulate the list
functions CONS,CAR,CDR,NIL as defined in
\cite{mccarthy60} (see proof in \cite{ppdp09pISO}).
\end{prop}

Using these predicates we define a bijection between
finite sequences represented as lists of their values and natural numbers
\begin{code}
list2nat([],0).
list2nat([X|Xs],N):-list2nat(Xs,N1),cons(X,N1,N).
\end{code}
\begin{code}
nat2list(0,[]).
nat2list(N,[X|Xs]):-N>0,hd(N,X),tl(N,T),nat2list(T,Xs).
\end{code}
working as follows:
{\small \begin{verbatim}
?- nat2list(2012,Ns),list2nat(Ns,N).
Ns = [2, 0, 0, 1, 0, 0, 0, 0],
N = 2012
\end{verbatim}}

\section{Ranking Hereditarily Finite Sequences} \label{hered}

\begin{df}
The {\em ranking problem} for a family of
combinatorial objects is finding a unique 
natural number associated to each object,
called its {\em rank}.
The inverse {\em unranking problem} consists of generating a unique
combinatorial object associated to each natural number. 
\end{df}

\begin{df}
A hereditarily finite sequence is {\tt []} or a finite sequence of
 hereditarily finite sequences.
\end{df}

We will describe, by instantiating the data type transformation described
in \cite{ppdp09pISO} how to extend a bijection
$\mathbb{N} \to  [\mathbb{N}]$ to trees representing
{\em hereditarily finite sequences}.
The two sides of the bijection are expressed as two higher order
predicates {\tt rank} and {\tt unrank}
parameterized by two transformations
{\tt F} and {\tt G}:
\begin{code}
unrank(F,N,Rs):-call(F,N,Ns),maplist(unrank(F),Ns,Rs).

rank(G,Ts,Rs):-maplist(rank(G),Ts,Xs),call(G,Xs,Rs).
\end{code}
These predicates can be seen as a form of ``structured recursion''
that propagate a simpler operation ({\tt F} and {\tt G}) guided by the structure of
the underlying data type. 
We can instantiate this mechanism to derive a bijection between natural
numbers and trees representing hereditarily finite sequences 
using {\tt rank} and {\tt unrank} as:
\begin{code}
nat2hfseq(N,T):-unrank(nat2list,N,T).

hfseq2nat(T,N):-rank(list2nat,T,N).
\end{code}
They work as follows:
{\small \begin{verbatim}
?- nat2hfseq(2012,HFSEQ),hfseq2nat(HFSEQ,N).
HFSEQ = [[[[]]], [], [], [[]], [], [], [], []],
N = 2012
\end{verbatim}}
One can represent the recursive {\em unfolding} of a natural number into
a hereditarily finite sequence as a directed ordered multigraph
(Fig. \ref{f2}). Note that as the mapping {\tt nat2list} generates
a sequence where the order of the edges matters, this order is
indicated with integers starting from {\tt 0} labeling the edges.
\FIG{f2}{2012 as a HFSEQ}{0.40}{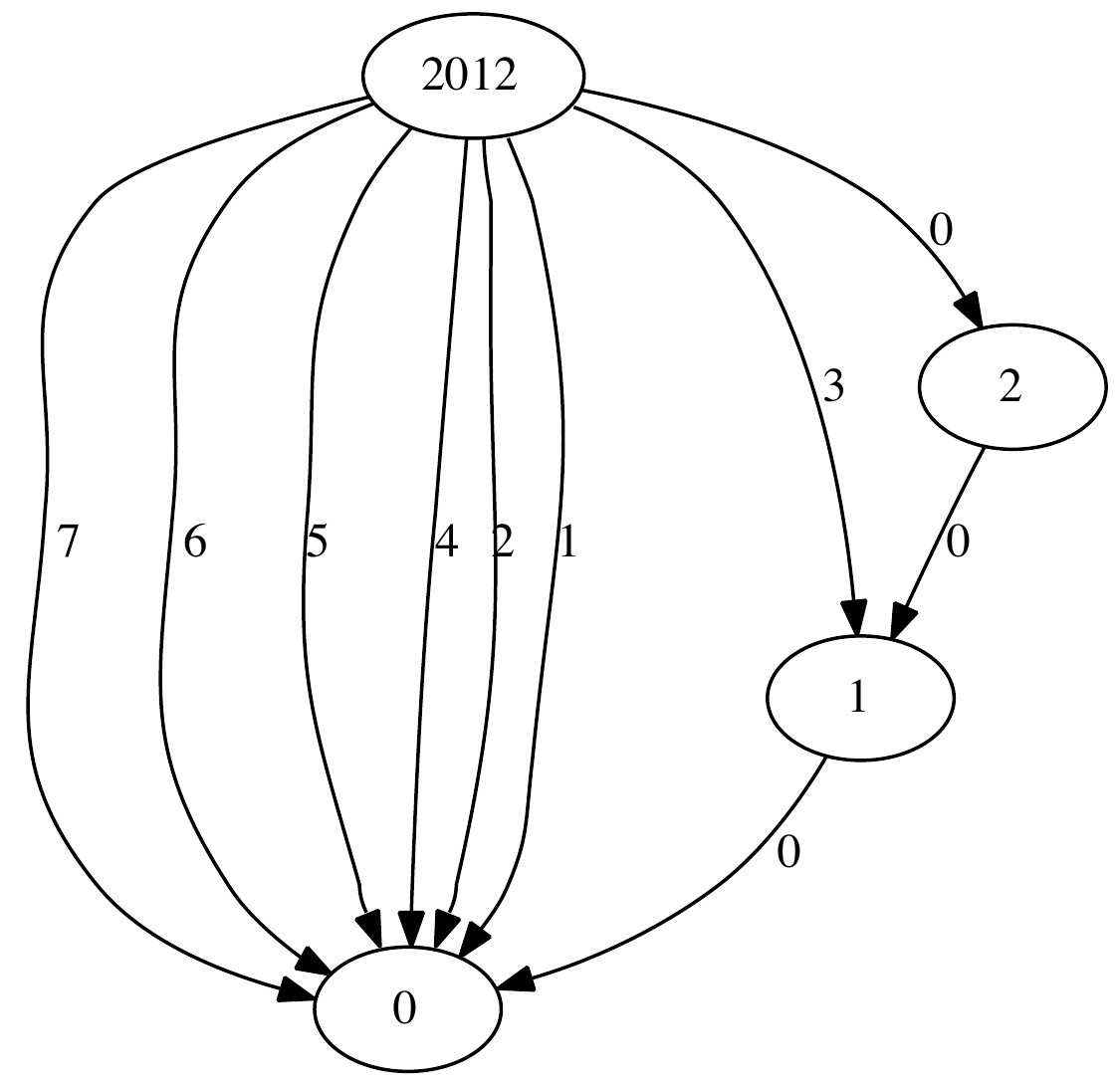}

\section{Computing with hereditarily finite sequences} \label{comput}
This section describes a surprising possibility
derived from the existence of bijections between various data
types and natural numbers. It answers positively the following question:
can we turn such bijections
into actual isomorphisms such that operations like
additions or multiplications defined on symbolic objects
(e.g. trees or parenthesis languages) mimic their
natural number equivalents? Moreover, we want a genuinely
constructive proof that this can be done, which means
that we need to build inductive definitions, starting
with successor and predecessor and then extend them
to implement everything else.

We will build these operations incrementally. We start with successor/predecessor 
operations and simple (but slow) mappings to natural numbers.
We then provide efficient implementations, working, like in the case of
bitstring representations, in time proportional to the size
of the operands.

\subsection{Successor and predecessor}
To derive efficient successor and predecessor operations we recall that
the equation {\tt Z=[X|Y]} on hereditarily finite sequences corresponds
bijectively to the equation 
\begin{equation}\label{nateq}
Z=2^X(2Y+1)
\end{equation}
on natural numbers. 
Successor and predecessor predicates {\tt s/2} and {\tt p/2}
are defined as:
\begin{code}
s([],[[]]).
s([[K|Ks]|Xs],[[],K1|Xs]):-p([K|Ks],K1).
s([[]|Xs],[[K1|Ks]|Ys]):-s(Xs,[K|Ys]),s(K,[K1|Ks]). 

p([[]],[]).
p([[],K|Xs],[[K1|Ks]|Xs]):-s(K,[K1|Ks]).
p([[K|Ks]|Xs],[[]|Zs]):-p([K|Ks],K1),p([K1|Xs],Zs).
\end{code}
The two predicates are deterministic and implement functions
when their first arguments are ground, given that the patterns 
used in the heads of the rules share no instances. 
If executed under a breadth-first evaluation rule (or if
impure Prolog operations are used) the two predicates can be merged 
into a single reversible predicate. We have preferred pure Horn clause
definitions, however, and reordered the goals in the clause bodies
as needed.

When navigating over hereditarily finite sequence trees,
{\tt s/2} implements tree transformations such that the following propositions hold:
\begin{prop}
If T is such that {\tt hfseq2nat(T,N)}, {\tt s(T,T1)} and {\tt hfseq2nat(T1,N1)} hold, then
N1 is N+1. 
\end{prop}
\begin{prop}
If {\tt T} (assumed different from []) is such that {\tt hfseq2nat(T,N)}, {\tt p(T,T1)} and {\tt hfseq2nat(T1,N1)} hold, then
N1 is N-1. 
\end{prop}  
One can rephrase this saying that the pair {\tt hfseq2nat} and {\tt nat2hfseq}
acts as a {\em isofunctor} that transports successor and predecessor operations
between natural numbers and hereditarily finite sequences.
A proof is obtained by structural induction on the first argument of the
two predicates after defining 
a mapping between a multiway tree type and a natural number type
supporting an axiomatization of Peano arithmetic.
Note that by replacing {\tt []} by {\tt 0} and each relation
of the form {\tt [X|Y]=Z} in the inductive
definition of {\tt s} and {\tt t} with
equations of the form $2^X*(2*Y+1)=Z$
one can obtain arithmetic
formulas that, after simplifications result in the
the usual arithmetic relations
defining {\tt s} and {\tt p}.

One can prove the correctness of {\tt s} and {\tt p} with respect to the
corresponding successor and predecessor operations 
on $\mathbb{N}$, by verifying that when interpreting each constructor
in terms of equation \ref{nateq} on $\mathbb{N}$ 
the resulting formulas become identities. 

For instance,
{\tt s([],[[]])} becomes $s(0,2^0*(2*0+1))$ and then $s(0,1)$ which states 
that the successor of {\tt 0} is {\tt 1}. 

On the other hand the second and third
recursive equations in the definitions of {\tt s} and {\tt p} become logical
implications between arithmetic identities, relatively easy to prove
through a sequence of simplifications.

For instance, the second equation in the definition of {\tt s/2} becomes,
after putting $[K|Ks]\rightarrow x$,$Xs \rightarrow y$, $K1 \rightarrow z$ with $x,y,z \in \mathbb{N}$.
\begin{equation}
s([x|y],[0,z|y]):-p(x,z).
\end{equation}
After interpreting {\tt :-} as inverse logical implication $\Leftarrow$ we obtain
\begin{equation}
s(2^x*(2*y+1),2^0*2*(2^z*(2*y+1))+1) \Leftarrow p(x,z).
\end{equation}
After interpreting {\tt s} and {\tt p} as successor and predecessor on
$\mathbb{N}$ we obtain:
\begin{equation}
1+(2^x*(2*y+1) = 2*2^z*(2*y+1)+1 \Leftarrow (x=z+1).
\end{equation}
After replacing $x$ by $z+1$ on the left side we obtain:
\begin{equation}
2^{z+1}*(2*y+1) = 2^{z+1}*(2*y+1)
\end{equation}
which is clearly an identity in $\mathbb{N}$.

Note that the ability to reason about the correctness
of our programs has been clearly facilitated by
the declarative semantics of Prolog, for instance
when interpreting {\tt :-} as reverse logical implication.

Let us now define, using {\tt s/2} and {\tt p/2} a simple (but inefficient) bijection
from trees (with leaves made of empty lists) 
to ordinary natural numbers:
\begin{code}
tree2nat([],0).
tree2nat([X|Xs],N):-p([X|Xs],Y),tree2nat(Y,M),N is M+1.

nat2tree(0,[]).
nat2tree(N,[X|Xs]):-N>0,M is N-1,nat2tree(M,Y),s(Y,[X|Xs]).
\end{code}
working as follows:
{\small \begin{verbatim}
?- nat2tree(2012,T),tree2nat(T,N).
T = [[[[]]], [], [], [[]], [], [], [], []],
N = 2012 ;
\end{verbatim}}
After defining a generator for the infinite stream of hereditarily finite sequences
mapped to successive natural numbers
\begin{code}
n([]).
n(S):-n(P),s(P,S).
\end{code}
one can confirm empirically that our two symbolic {\tt s/2} and {\tt p/2} operations provide
indeed emulations of their standard counterparts:
{\small \begin{verbatim}
?- n(X),tree2nat(X,N).
X = [], N = 0 ;
X = [[]], N = 1 ;
X = [[[]]], N = 2 ;
X = [[], []], N = 3 ;
.......
\end{verbatim}}

\subsection{Simple arithmetic operations in terms of successor and predecessor}
The {\tt s/2} and {\tt p/2} predicate pair can be used to
implement the usual arithmetic operations
in time {\tt O(N)} where {\tt N} is the natural number corresponding
to the first operand. For instance,
addition can be defined as follows:
\begin{code}
slow_add([],X,X).
slow_add([X|Xs],Y,Z):-p([X|Xs],P),s(Y,Y1),slow_add(P,Y1,Z).
\end{code}
It works indeed as expected:
{\small \begin{verbatim}
?- nat2tree(42,T),slow_add(T,T,R),tree2nat(R,N).
T = [[[]], [[]], [[]]], R = [[[[]]], [[]], [[]]], N = 84
\end{verbatim}}
We will next define efficient operations, with asymptotic complexity
comparable to typical bignum packages provided by various languages.

\subsection{Basic recognizers and constructors}
We start with recognizers
for odd numbers {\tt o\_/2}, strictly positive even numbers {\tt i\_/2} and zero
{\tt e\_/1}.
\begin{code}
o_([[]|_]).
i_([[_|_]|_]).
e_([]).
\end{code}
Next, we define our constructors. 
The first one, {\tt o/2} builds odd numbers,
 as if provided by leftshift+increment operation
{\tt 2*X+1}
 The later applies the successor predicate to the result of the first,
as if provided by the {\tt 2*X+2} operation. 
\begin{code}
o(X,[[]|X]).
i(X,Y):-s([[]|X],Y).
\end{code}
Note that the predicate {\tt e\_/1}
can also be seen as a constructor for the empty list representing {\tt 0}.

\subsection{Arithmetic operations with hereditarily finite sequences -- efficiently}
To provide efficient, possibly practical implementations of
arithmetic operations,
we will need a few more steps towards emulating binary representations
including variants of left and right shifting operations.

\subsubsection{Deconstructing}
Let us first build a deconstructor {\tt r/2}, working as a 
decrement + rightshift operation on bitstrings such that it maps
both {\tt 2*X+1} and {\tt 2*X+2} to {\tt X}, i.e. such that it
reverses the action of the constructors
{\tt o/2} and {\tt i/2}.

\begin{code} 
r([[]|Xs],Xs).
r([[X|Xs]|Ys],Rs):-p([[X|Xs]|Ys],[[]|Rs]).
\end{code}
Note that the first clause maps to {\tt n} a term corresponding to an odd number of the
form {\tt 2*n+1}, while the second applies the predecessor to an even
number while trimming the result (an odd number) in a similar way to the
first clause.

\subsubsection{Converting back and forth}
Given the deconstructor {\tt r/2} and the constructors {\tt o/2} and {\tt i/2},
we can empirically validate the intuitions behind our symbolic representations,
by mapping them one-to-one to conventional natural numbers.

We first define a converter {\tt s2n/2}, mapping tree representations of
hereditarily finite sequences to conventional natural numbers:
\begin{code}
s2n([],0).
s2n(X,R):-o_(X),r(X,S),s2n(S,N),R is 1+2*N.
s2n(X,R):-i_(X),r(X,S),s2n(S,N),R is 2+2*N.
\end{code}
then a converter {\tt n2s/2} from natural numbers to our symbolic representations:
\begin{code}
n2s(0,[]).
n2s(N,R):-N>0,P is N mod 2,N1 is (N-1) // 2,
  n2s(N1,X),
  ( P=:=0->i(X,R)
  ; o(X,R)
  ).
\end{code}
They work as expected, and {\tt s2n} can be seen as enumerating
the stream of natural numbers correctly.
{\small \begin{verbatim}
?- n2s(42,S),s2n(S,N).
S = [[[]], [[]], [[]]],
N = 42

?-n(X),s2n(X,N).
X = [], N = 0 ;
X = [[]], N = 1 ;
X = [[[]]], N = 2 ;
X = [[], []], N = 3 ;
.......
\end{verbatim}}
Note also that they work in time proportional to the size of the
representations.

\subsubsection{Efficient Addition}
Guided by this mapping, that sees our symbolic representations as if they
were bitstrings in {\em bijective base-2}, we can implement an addition
operation working in time proportional to the size of the operands:
\begin{code}
a([],Y,Y).
a([X|Xs],[],[X|Xs]).
a(X,Y,Z):-o_(X),o_(Y),a1(X,Y,R),  i(R,Z).
a(X,Y,Z):-o_(X),i_(Y),a1(X,Y,R), a2(R,Z).
a(X,Y,Z):-i_(X),o_(Y),a1(X,Y,R), a2(R,Z).
a(X,Y,Z):-i_(X),i_(Y),a1(X,Y,R), s(R,S),i(S,Z).

  a1(X,Y,R):-r(X,RX),r(Y,RY),a(RX,RY,R).
  a2(R,Z):-s(R,S),o(S,Z).
\end{code}
working instantly on arbitrarily large natural numbers:
{\small \begin{verbatim}
?-n2s(12345678901234567890,A),n2s(10000000000000000000,B),a(A,B,S),s2n(S,N).
A = [[[]], [[[]]], [[]], [], [[]], [[]], [[[...]]], [], []|...],
B = [[[], [], [[[]]]], [[]], [], [], [], [[[]]], [[], []], [[...]]...|...],
S = [[[]], [[[]]], [[]], [], [[]], [[]], [[[...]]], [], []|...],
N = 22345678901234567890 .
\end{verbatim}}

\subsubsection{Efficient Multiplication}
We can implement efficient multiplication guided by intuitions about
binary multiplication in base {\tt 2} and bijective-base {\tt 2} as follows:
\begin{code}
m([],_,[]).
m(_,[],[]).
m(X,Y,Z):-p(X,X1),p(Y,Y1),m0(X1,Y1,Z1),s(Z1,Z).
  
m0([],Y,Y).
m0([[]|X],Y,[[]|Z]):- m0(X,Y,Z).
m0(X,Y, Z):-i_(X),r(X,X1),m0(X1,Y,Z1),a(Y,[[]|Z1],Y1),s(Y1,Z).
\end{code}
One can see that it handles easily large numbers (the {\em googol}$=10^{100}$ included!):
{\small \begin{verbatim}
?-n2s(12345678901234567890,A),n2s(10000000000000000000,B),m(A,B,S),s2n(S,N).
A = [[[]], [[[]]], [[]], [], [[]], [[]], [[[...]]], [], []|...],
B = [[[], [], [[[]]]], [[]], [], [], [], [[[]]], [[], []], [[...]]|...],
S = [[[[[]]], [[]]], [[]], [[], [[]]], [], [[[[]]]], [[]], []|...],
N = 123456789012345678900000000000000000000 .

?- n2s((10^100),Googol),m(Googol,Googol,S),s2n(S,N).
Googol = [[[[[]]], [[[]]], []], [[], []], [], [], [], [[], []], [[]] |...],
S = [[[[], []], [[[]]], []], [[[[]]]], [], [], [[]], [[[]]], [] |...],
N = 100000000................00000000000000000000000000000000
\end{verbatim}}
Let $<\mathbb{T},a,m>$ denote the algebraic structure induced by
the operations {\tt a} and {\tt m} on the set of
multiway trees representing hereditarily finite sequences
and $<\mathbb{N},+,*>$ the corresponding algebraic structure
on natural numbers with addition and multiplication.
The following holds:
\begin{prop}
The addition and multiplication operations {\tt a/3} and {\tt m/3}
induce an isomorphism between the semirings with commutative
multiplication $<\mathbb{N},+,*>$ and $<\mathbb{T},a,m>$.
\end{prop}

We conclude this first part of the paper by confessing that inventing 
(the asymptotically efficient)
Horn clause definitions of various arithmetic operations would not have
been possible without the ``reverse engineering" capabilities
provided by the data transformation framework in \cite{ppdp09pISO},
which has enabled us to move at will between representations like
bijective base-2 binary numbers, bit-stacks, hereditarily finite sets,
hereditarily finite sequences and watch the internal workings
of ordinary operations through functors defined between
these domains. 

While page limits do not allow us to describe this process
in full detail, we have extended these operations to cover, with
asymptotic complexity comparable to standard bignum packages, to comparaisons,
subtraction, division, powers etc.

\section{Computing with binary trees representing G\"odel's System {\bf T} types} \label{systemT}

\begin{df}
In  G\"odel's System {\bf T} \cite{goedel1958bisher}
a type is either $N$ or $t \rightarrow s$ where $t$ and $s$ are types.
\end{df}
The basic type $N$ usually stands for the type of natural numbers.
We will briefly show here that natural numbers can be emulated
directly with types, by using a single constant $e$ as basic type,
representing $0$.

First, we observe that, guided by the known isomorphism between 
multiway and binary trees\footnote{That manifests itself in languages like
Prolog or
LISP as the dual view of lists as a representation of sequences or binary
CONS-cell trees.},
we can bring with a {\em functor} defined from hereditarily finite sequences to
binary trees the definitions of {\tt s/2} and {\tt p/2} into
corresponding definitions in the language of system {\bf T} 
types, {\tt s\_/2} and {\tt p\_/2}.

\begin{code}
s_(e, (e->e)).
s_(((K->Ks)->Xs), (e->(K1->Xs))) :- p_((K->Ks), K1).
s_((e->Xs), ((K1->Ks)->Ys)) :- s_(Xs, (K->Ys)), s_(K, (K1->Ks)). 
\end{code}
\begin{code}
p_((e->e), e).
p_((e->(K->Xs)), ((K1->Ks)->Xs)) :- s_(K, (K1->Ks)).
p_(((K->Ks)->Xs), (e->Zs)) :- p_((K->Ks), K1), p_((K1->Xs), Zs).
\end{code}

The following example illustrates that {\tt s\_} and {\tt p\_} work as expected:
{\small \begin{verbatim}
?- s_(e,One),s_(One,Two),s_(Two,Three),s_(Three,Four),p_(Four,Three).
One = (e->e),
Two = ((e->e)->e),
Three = (e->e->e),
Four = (((e->e)->e)->e)
\end{verbatim}}

We will only give here the code of a generator {\tt
n\_/1}  for the infinite stream of natural numbers
represented as types in system {\bf T}, and a simple
converter to usual natural
numbers {\tt t2n}, modeled after {\tt tree2nat/2}.
\begin{code}
n_(e).
n_(S):-n_(P),s_(P,S).

t2n(e,0).
t2n((T->S),N):-p_((T->S),U),t2n(U,M),N is M+1.
\end{code}
confirming empirically that our computations mimic the usual ones:
{\small \begin{verbatim}
?-  n_(T),t2n(T,N).
T = e, N = 0 ;
T = (e->e), N = 1 ;
T = ((e->e)->e), N = 2 ;
T = (e->e->e), N = 3 ;
T = (((e->e)->e)->e), N = 4 ;
...
\end{verbatim}}
Fast arithmetic computations, operating directly on types,
can be derived using the corresponding code for hereditarily
finite sequences as ``boilerplate''.

\paragraph{Deriving a bidirectional successor/predecessor predicate}
The predicates {\tt s\_} and {\tt p\_} are mutually recursive and
structurally similar. Moreover, each of them would run reversibly under
a breadth-first evaluation order. An interesting challenge is to
derive a bidirectional variant replacing both predicates.
One could achieve this by using impure operations like
{\tt nonvar/1} to check which argument is instantiated or, equivalently,
checking the instantiation of the arguments using negation as
failure.
We proceed by merging the two predicates' shared clauses and
adding an extra argument taking the values {\tt up} or {\tt down}
to indicate which way the the computation goes.
\begin{codex}
sp(e, (e->e), _).
sp(((K->Ks)->Xs), (e->(K1->Xs)),Dir) :-
  flip(Dir,Other), 
  sp(K1,(K->Ks), Other).
sp((e->Xs), ((K1->Ks)->Ys), up) :- 
  sp(Xs, (K->Ys) ,up), 
  sp(K, (K1->Ks), up). 
\end{codex}
\begin{codex}
sp((e->Xs), ((K1->Ks)->Ys), down) :- 
  sp(K, (K1->Ks), down), 
  sp(Xs, (K->Ys) ,down).
\end{codex}
\begin{codex}
flip(up,down).
flip(down,up).

up_or_down(_X,Y,down):- \+(Y=other).
up_or_down(X,_Y,up):- \+(X=other).

sp(X,Y):-up_or_down(X,Y,Dir),sp(X,Y,Dir).
\end{codex}
Note also the auxiliary predicate {\tt flip/2}, which indicates a change of direction,
and the auxiliary predicate {\tt up\_or\_down}, that choses among the two possible
directions, based on the instantiation of at least one of the arguments of {\tt sp/2}.
We detect instantiation of the arguments
testing them against the atom {\tt other}, assumed not to be part of 
the Herbrand Universe of our program.

One step further, we push the call to {\tt sp/3} into {\tt flip/2}
(as it is the only continuation of {\tt flip/2}), and merge the last two clauses,
while delegating the ordering of the recursive calls to 
the auxiliary predicate {\tt order\_sp}. Note that we also fold 
{\tt up\_or\_down} as part of the definition of {\tt sp/2}.
\begin{code}
sp(e, (e->e), _).
sp(((K->Ks)->Xs), (e->(K1->Xs)), Dir):-flip_sp(Dir, K1, (K->Ks)).
sp((e->Xs), ((K1->Ks)->Ys), Dir):-order_sp(Dir, Xs, (K->Ys), K, (K1->Ks)). 

flip_sp(up,X,Y) :- sp(X,Y,down).
flip_sp(down,X,Y) :- sp(X,Y,up).

order_sp(up,A,B,C,D) :- sp(A,B,up), sp(C,D,up).
order_sp(down,A,B,C,D) :- sp(C,D,down), sp(A,B,down).

sp(X,Y) :- \+(X=other), sp(X,Y,up).
sp(X,Y) :- \+(Y=other), sp(X,Y,down).
\end{code}
One can try out {\tt sp/2} working as a bidirectional 
successor/predecessor predicate when at least one of its arguments 
is instantiated:
{\small \begin{verbatim}
?- sp(Pred,((e->e)->e)). 
Pred = (e->e) .

?- sp((e->e),Succ).
Succ = ((e->e)->e) .

?- sp((e->e),((e -> e) -> e)).
true.
\end{verbatim}}

\section{Mapping hereditarily finite sequences to parenthesis
languages} \label{parlang}

We will next explore the bijection between hereditarily finite sequences
and the language of balanced parenthesis, known to combinatorialists \cite{berstel2002formal,liebehenschel2000ranking,bertoni2006context}
as a member of the {\em Catalan family}, which
also includes the binary trees representing {\tt System T} types.

An encoder for the balanced parenthesis language is obtained by
combining a parser and a writer, which, with some ingenuity, can be made
one and the same in a language like Prolog. 

As hereditarily finite
sequences naturally map one-to-one to parenthesis expressions expressed as
bitstrings, we will choose them as target of the transformers.
Our parser recurses over a bitstring (encoding balanced parentheses {\tt '['} as {\tt 0},
{\tt ']'} as {\tt 1}) and builds a {\tt HFSEQ} tree {\tt T}:
\begin{code}
pars_hfseq(Xs,T):-pars2term(0,1,T,Xs,[]).

pars2term(L,R,Xs) --> [L],pars2args(L,R,Xs).

pars2args(_,R,[]) --> [R].
pars2args(L,R,[X|Xs])-->pars2term(L,R,X),pars2args(L,R,Xs).   
\end{code}
Note that {\tt pars\_hfseq} is {\em bidirectional} i.e. it works both as an
encoder and a decoder:
{\small \begin{verbatim}
?- pars_hfseq([0,0,1,0,1,1],T),pars_hfseq(Ps,T).
T = [[], []],
Ps = [0, 0, 1, 0, 1, 1]
\end{verbatim}}
One can see the bijection defined by {\tt pars\_hfseq} as 
a bridge between a family of formal
languages and  
hereditarily finite sequences, represented as multiway trees.


\paragraph{Kraft's inequality}
As the sequences computed by {\tt pars\_hfseq} are elements of
the balanced parenthesis language (also called Dyck primes)
\cite{berstel2002balanced}, they implement {\em uniquely decodable} {\em
self-delimiting} codes.
Moreover, each of them is also a {\em prefix code}, i.e. there's no
way to add a string made of any combination of balanced
left or right parenthesis at the end of a code and obtain another
code. For a similar reason, each of them is also a {\em suffix code}.
Such codes are known in the literature under a variety of different names i.e.
as {\em reversible variable-length codes}, {\em bifix codes}
or {\em fix-free} codes\footnote{A nice property of such codes is
that parallel bidirectional decoding is possible.
Also, the ability to decode from either the beginning or the end
makes them suitable for encoding media streams.}.

In particular, given that they are {\em uniquely decodable} codes,
it follows that the {\em Kraft inequality} 
\cite{kraft1949device}
holds for them, i.e. if $l_0,l_1 \ldots l_k \ldots$ denote the length of the codes, then
\begin{equation}
\sum_{k \geq 0} 2^{-l_k} \leq 1 
\end{equation}
We define the function computing the left side of the 
{\em Kraft} inequality (called {\em Kraft-sum}), 
and the corresponding test as follows.
\begin{code}
kraft_sum(M,S):- M1 is M-1, numlist(0,M1,Ns),
  maplist(kraft_term,Ns,Ls),  
  sumlist(Ls,S).

kraft_term(N,X):-parsize(N,L), X is 1/2^L.

parsize(N,L):- nat2hfseq(N,HFSEQ), pars_hfseq(Xs,HFSEQ), length(Xs,L).

kraft_inequality(M):-kraft_sum(M,S),S=<1.
\end{code}
The following example illustrates that the Kraft's inequality holds and it is likely that the Kraft-sum converges to a value below 0.5:
{\small \begin{verbatim}
?- maplist(kraft_sum,[10,100,1000,2000,3000,4000],R).
R = [0.364258, 0.382935, 0.390383, 0.391615, 0.392292, 0.392598]
\end{verbatim}}

The bijection between hereditarily finite sequences and balanced parenthesis
languages provides a succinct alternative representation for purposes
of efficient arithmetic operations using bitvector operations -- by encoding
the two parenthesis as {\tt 0} and {\tt 1}.
As a possible practical application, this allows building in Prolog, at source level,
a library supporting arbitrary length arithmetic operations.

\section{Related work} \label{related}
{\em Ranking} functions can be traced back to G\"{o}del numberings
\cite{Goedel:31,conf/icalp/HartmanisB74} associated to formulae. 
Together with their inverse {\em unranking} functions they are also 
used in combinatorial generation
algorithms \cite{conf/mfcs/MartinezM03,knuth_boolean}.
Natural number encodings of hereditarily finite sets have 
triggered the interest of researchers in fields ranging from 
Axiomatic Set Theory and Foundations of Logic to 
Complexity Theory and Combinatorics
\cite{finitemath,kaye07,DBLP:journals/mlq/Kirby07}.

The encodings of hereditarily finite sets and sequences described in this paper
originate in \cite{ppdp09pISO,calc09fiso,sac09fISO,arxiv:fISO}. The key difference
is that while in our previous work we use pairs of bijections
encapsulated as higher order predicates/functions to define various isomorphisms
directly, here we provide actual algorithms for arithmetic
operations, ordering etc. while in our previous work the
existence of such algorithms was only implied ``non-constructively".

An emulation of Peano and conventional binary arithmetic operations
in Prolog, is described in \cite{DBLP:conf/flops/KiselyovBFS08}.
Their approach is similar as far as a symbolic representation is used.
The key difference with this paper is that our operations work on tree
structures, and as such, they are not based on previously known algorithms.
Our tree-based algorithms are also likely to support
parallel execution in a way similar to the 
powerlists of \cite{Misra94powerlist:a}.
Arithmetic computations with types expressed as {\tt C++} templates
are described in \cite{DBLP:journals/corr/cs-CL-0104010} and in
online articles by Oleg Kiselyov using Haskell's
type inference mechanism. However, the mechanism advocated there is
basically the same as \cite{DBLP:conf/flops/KiselyovBFS08}, focusing
on Peano and binary arithmetics.
The connection between hereditarily finite sequences
and balanced parenthesis languages places them the context of the
well known to combinatorialists {\em Catalan families}
\cite{berstel2002formal,liebehenschel2000ranking,bertoni2006context}.

\section{Conclusion} \label{concl}

We have derived a few algorithms expressing arithmetic
computations symbolically, in terms of
hereditarily finite sequences and types in G\"odel's
system {\bf T}. 

This has been made possible by
extending the techniques introduced in \cite{ppdp09pISO} that
allow observing the internal working of intricate
mathematical concepts through isomorphisms transporting
operations between fundamental data types.

At the same time, we have shown that logic programming provides
a flexible framework for modeling  mathematical concepts
from fields as diverse as combinatorics, formal languages,
type theory and coding theory.

Arithmetic operations with hereditarily finite sequences
are likely to be interesting for hardware (FPGA) implementations of
large integer operations used in cryptography. They are also subject to 
parallelization \cite{Misra94powerlist:a} and can
provide computations with giant numbers that do not
fit in any computer memory with a flat bitstring representation\footnote{
Something as simple as {\tt [[[[[[[[[]]]]]]]]]} expresses a very large number -
as such numbers correspond to towers of exponents of the form $2^{.^{.2^{2}}}$.}.

Reversible variable length (bifix) codes like the ones we
derived in section \ref{parlang} have found uses in
image and video coding \cite{wen1998reversible} (including MPEG4!).
Prefix codes are used in defining modern versions
of Kolmogorov complexity \cite{vitanyi}. The fact
that this property holds, recursively, for arbitrary
parts of the code, combined with their {\em ability to express
programming language constructs}, as shown in \cite{ppdp09pISO},
makes them an interesting alternative to the Elias
codes \cite{elias75} typically used in the field.

\section*{Acknowledgment} We thank NSF (research grant 1018172) for support.

\bibliographystyle{INCLUDES/splncs}
\bibliography{tarau,INCLUDES/theory,INCLUDES/proglang,INCLUDES/biblio,INCLUDES/syn}

\end{document}